\documentclass{mn2e}
\usepackage[cp1251]{inputenc}
\usepackage{graphicx,epsfig}
\usepackage{floatflt}
\usepackage {amssymb}
\usepackage {amsmath}
\usepackage {longtable}
\usepackage[usenames]{color}
\title[Three Einstein rings]
    {Three Einstein rings: explicit solution and numerical simulation}
\author[E.Yu. Bannikova and A.T. Kotvytskiy]
  {E. Yu.~Bannikova$^{1,2}$\thanks {E-mail: bannikova@astron.kharkov.ua} and A. T.~Kotvytskiy$^{1,2}$\\
   $1$   Institute of Radio Astronomy of Nat.Ac.Sci. of Ukraine, Krasnoznamennaya 4,
   61022 Kharkov, Ukraine \\
   $^2$  Karazin Kharkov National University,
   Svobody Sq. 4, 61022 Kharkov, Ukraine \\
   }

\date{Accepted 2014 October 1.  Received 2014 September 11; in
original form 2014 July 11}

\pagerange{\pageref{firstpage}--\pageref{lastpage}} \pubyear{2014}

\def\LaTeX{L\kern-.36em\raise.3ex\hbox{a}\kern-.15em
    T\kern-.1667em\lower.7ex\hbox{E}\kern-.125emX}

\begin{document}

\label{firstpage}

\maketitle

\begin{abstract}
We investigated the effects of gravitational lensing for a system in
which a lens is a point mass and a homogeneous disc with a central
hole. In such system there is a variety of cases
resulting in formation of one, two and three Einstein rings. We
found an explicit solution and considered conditions for existence
of the second Einstein ring arising on the disc. Numerical modelling
of the images was made for various ratios of the central mass to the
disc one and for various values of the disc surface density. We also
analysed dependence of the magnification factor on a source position
for such system. The result of our work can be used in search of
astrophysical objects with a toroidal (ring) structure.
\end{abstract}


\section{INTRODUCTION}
The gravitational lensing phenomenon is presently believed to be an extremely powerful
tool to investigate a spatial distribution of matter at different scales in the Universe. The fundamentals of gravitational lensing are well developed and can be found in many books and reviews, such as, e.g., (Bliokh \& Minakov 1989; Schneider, Ehlers \& Falco 1999; Refsdal \& Surdej 1994; Zakharov 1997; Schneider, Kochanek \& Wambsganss 2006). Observations have already provided a large gallery of gravitational lenses, and every year is adding several newly discovered objects varying in their features.

The most spectacular effect of gravitational lensing is the Einstein
ring, which arises when a point lens happens to be at the line of
sight to the source (Einstein  1936). In more complex lens systems,
multiple Einstein rings may arise. For example, gravitational
lensing by a black hole leads to the formation of multiple Einstein
rings (Ohanian 1987; Virbhadra \& Ellis 2000; Virbhadra 2009; Bisnovatyi-Kogan \&
Tsupko 2012).
Such exotic lens as naked singularity in the scalar field can lead 
to formation of two Einstein rings (Virbhadra, Narasimha \& Chitre 1998; Virbhadra \& Ellis 2002).
 In the other case, two Einstein rings can be formed due to
lensing by two point masses located at two different distances along
the line of sight ”source-observer”. If a more distant lens is
luminous, it may play a role of the second source thus leading to
formation of three Einstein rings (Werner et al. 2008). Recently, a
lens system with a partial double Einstein ring was discovered –
SDSSJ0946+1006 (Gavazzi et al. 2008). The authors suggested that
these Einstein rings are formed from two sources situated at
different distances from the lens.

Lensing by a ring (without a central mass) was considered in short
in the book by Schneider, Ehlers \& Falco (1999), where a
possibility of formation of the second ring was mentioned, with a
notation that the problem of lensing by a ring structure is of
”little interest in astrophysics”.  However, this problem acquires much more importance if a central mass is contained in this lens system.
 Indeed, the presence of a central mass is a necessary
 condition for stability of a self-gravitating torus or ring
 (Bannikova, Vakulik \& Sergeev 2012).

 The collision of galaxies can lead to the formation of a ring
around the central dense core (Gerber, Lamb  Balsara, 1996). Such
galaxies are known as ring galaxies (Athanassoula \& Bosma, 1985)
and typical representative of them is the Hoag’s object in which the
ring-like distribution of matter around the central galaxy has a
perfectly circular shape (Hoag 1950; Schweizer et al. 1987). It is
likely that ring structures can be formed in a dark matter.  In the
galaxy cluster C10024+17, the ring structure has been found in
distribution of dark matter with the use of gravitational lensing
method (Jee et al. 2007). It was assumed that such ring structure
was the result of a collision between two clusters. We may assume
that the ring structures of dark matter can also occur in collisions
of not only clusters but of galaxies too. Therefore, it is  of
interest to investigate the effects of gravitational lensing by such
systems.

In this paper, we consider the case where the lens is a central
point mass and a ring (hereinafter a disc with a hole) with a
homogeneous density distribution. In Section 2 we obtain lens
equation and analytical expressions for the Einstein ring radii in
the case of a point source.  In Section 3 we obtain expressions for
the inner and outer boundaries of the Einstein rings (for finite
source). In Section 4 we investigate conditions for  existence of
the second Einstein ring and analyse various cases in which one, two
or three Einstein rings are formed. Since the considered case is a
generalization of the known cases of lensing by simpler systems, we
analyse limiting passage (Section 5). In Section 6, we consider a
magnification factor for different parameters of the lensing system.
In many cases, we present the images obtained numerically not only
for a source located on the line of sight, but also for
sources that are offset from it.

\section{LENS EQUATION FOR A SYSTEM CONSISTED OF A HOMOGENEOUS DISC AND A CENTRAL MASS}

We consider a thin annular disc with inner ($R_1$) and outer ($R_2$) radii
and a point mass located in the centre of symmetry. The total mass
of the system is a sum of the disc mass ($M_{disc}$) and the
central mass ($M_0$): $M = M_0 + M_{disc}$. Let us denote  $\boldsymbol{\eta}$ and
$\boldsymbol{\xi}$ as coordinate vectors of a source and an image, respectively.
The lens equation has the form (Schneider, Ehlers \& Falco 1999):
\begin{equation}\label{eq1}
\boldsymbol{y} = \boldsymbol{x} - \boldsymbol{\alpha}(\boldsymbol{x}),
\end{equation}
where $\boldsymbol{y} = \left(D_d/D_s\right)\left(\boldsymbol{\eta}/\xi_0\right)$ and $\boldsymbol{x} = \boldsymbol{\xi}/\xi_0$ are the dimensionless coordinates of the source and image. The radius of the Einstein ring for the total mass is
\begin{equation}\label{eq2}
\xi_0 = \sqrt{\frac{4GM}{c^2}\cdot \frac{D_{ds}D_d}{D_s}} ,
\end{equation}
$D_s$ and $D_d$ are distances from observer to the source and lens, respectively; $D_{ds}$ is a distance between the lens and the source. Since the lens is a system consisting of the central mass and the disc, the total deflection angle is
\begin{equation}\label{eq3}
\boldsymbol{\alpha} = \boldsymbol{\alpha}_0 + \boldsymbol{\alpha}_{disc},
\end{equation}
where the deflection angle due to the central mass is
\begin{equation}\label{eq4}
\boldsymbol{\alpha}_0 = m_0\frac{\boldsymbol{x}}{x^2},
\end{equation}
$m_0 = M_0/M$ is the central mass normalized to the total mass of the system.
The angle related to the light deflection due to the gravitational influence of the disc with an arbitrary density distribution has the form
\begin{equation}\label{eq5}
\boldsymbol{\alpha}_{disc} = \frac{1}{\pi}\int d^2 x' \kappa (\boldsymbol{x}')
                            \frac{\boldsymbol{x}-\boldsymbol{x}'}{\lvert \boldsymbol{x}-\boldsymbol{x}' \rvert^2}.
\end{equation}
Here the relative mass of the disc is $m_{disc} = M_{disc}/M$;
$\boldsymbol{x}'$ is the vector up to a square element of the disc
and the integration is over  total square of the disc. Relative
surface density\footnote{We factored $1/\pi$ out the integral
because this factor vanishes after integration over azimuthal angle.
So, the number $\pi$  is absent in the expression for the surface
density.} for an arbitrary law of distribution can be
represented as $\kappa(\boldsymbol{x}) = \kappa_0
f(\boldsymbol{x})$, where $\kappa_0 = m_{disc}/(r_2^2 - r_1^2)$ and
$r_{1,2} = R_{1,2}/\xi_0$. In the case of a homogeneous density
distribution $\kappa = Const=\kappa_0$.
\begin{figure}\label{Fig1}
\includegraphics[width = 80mm]{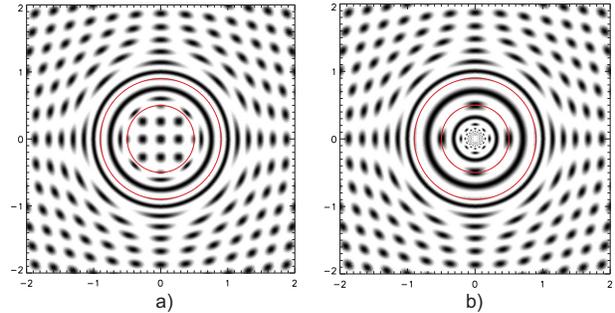}
\caption{Images of the grid of Gaussian sources (15x15), each of the
radius  $r_s=0.1$, formed in gravitational lensing by a) only a
homogeneous annular disc $m_{disc}=1$; b)  the central mass $m_0=0.1$ and
the disc. In both figures the inner radius $r_1=0.5$ and the outer
radius $r_2=0.9$. In all following figures the boundaries of annular lens are
shown by the thin red lines.}
\end{figure}
\begin{figure}\label{Fig2}
\includegraphics[width = 80mm]{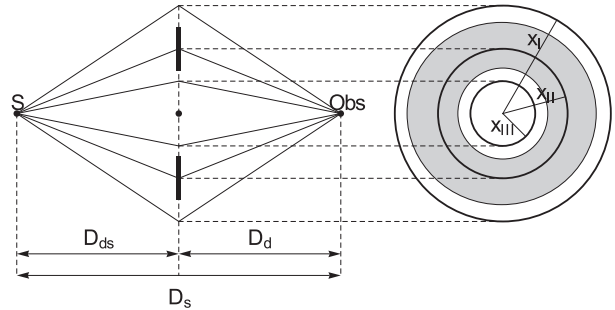}
\caption{The scheme of formation of the Einstein rings.}
\end{figure}
So, the lens equation (\ref{eq1}) for the central mass and the disc
has the following form
\begin{equation}\label{eq6}
\boldsymbol{y} = \boldsymbol{x} - m_0 \frac{\boldsymbol{x}}{x^2} -
\frac{1}{\pi}\int d^2 x' \kappa (\boldsymbol{x}')
                            \frac{\boldsymbol{x}-\boldsymbol{x}'}{\lvert \boldsymbol{x}-\boldsymbol{x}' \rvert^2},
\end{equation}
where normalized masses satisfy condition $m_0 + m_{disc} = 1$.
\begin{figure*}\label{Fig3}
\includegraphics[width = 150mm]{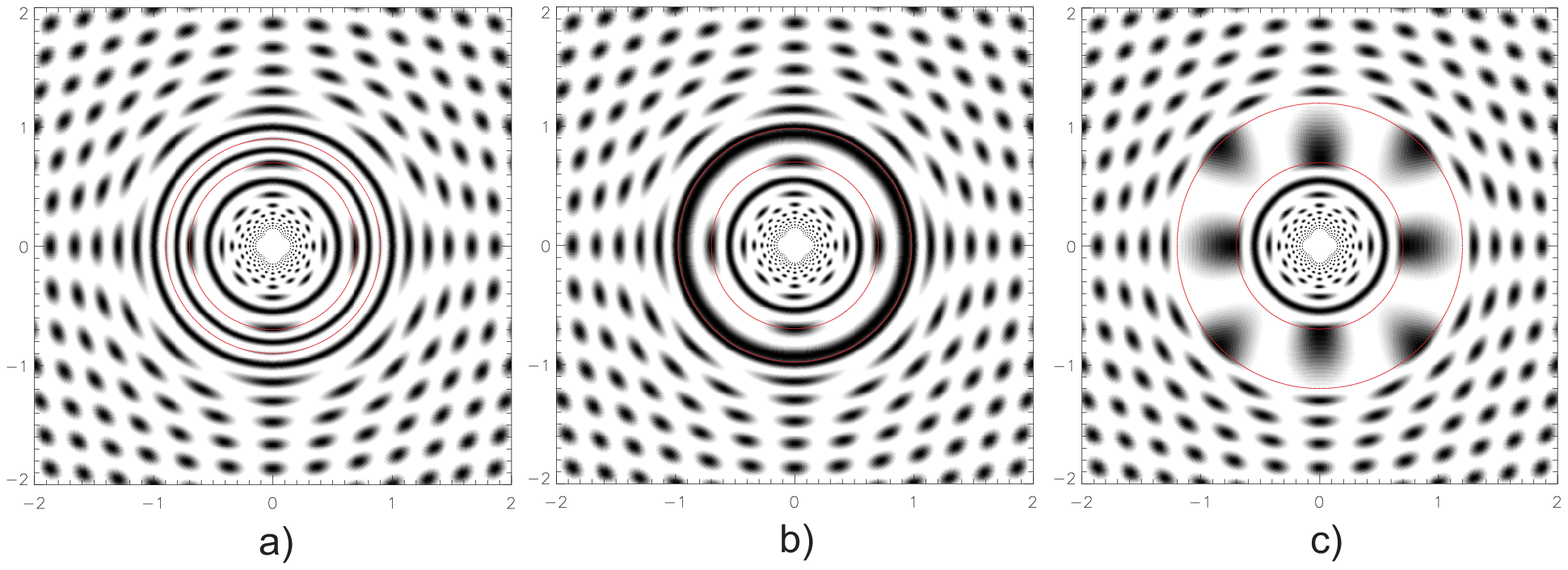}
\caption{Images of the grid of Gaussian sources ($r_s=0.1$) due to
gravitational lensing by a central point mass with $m_0=0.3$ and
a homogeneous disc with the inner radius $r_1=0.7$, for three
cases: a) $r_2=0.9$, b)$r_2=0.98$, c)$r_2=1.2$.}
\end{figure*}
Since the considered system is axially symmetric we pass to the
polar system of coordinates: $\boldsymbol{x}' = x'\times
(\cos\varphi, \sin\varphi)$ and $\kappa(\boldsymbol{x}') =
\kappa(x')$. After integration over azimuthal angle $\varphi$
(see Zakharov 1997), the lens equation (\ref{eq6})
transforms to three vector equations
\begin{equation}\label{eq7}
\boldsymbol{y} = \boldsymbol{x} \times
\begin{cases}
\left( 1 - \dfrac{m_0}{x^2}
-\dfrac{2}{x^2}\int\limits_{r_1}\limits^{r_2}
\kappa(x')x'dx'\right),
                & \mbox{$x \geq r_2$} 
                \\
\left( 1 - \dfrac{m_0}{x^2}
-\dfrac{2}{x^2}\int\limits_{r_1}\limits^{x} \kappa(x')x'dx'\right),
& \mbox{$r_1\leq x \leq r_2$}
                \\
\left( 1 - \dfrac{m_0}{x^2}\right), & \mbox{$ x \leq r_1$}.
\end{cases}
\end{equation}
It is easy to see that in the case of homogeneous density
distribution the equation  (\ref{eq7}) takes the form
\begin{equation}\label{eq8}
\boldsymbol{y} = \boldsymbol{x} \times
\begin{cases}
\left( 1 - \dfrac{1}{x^2} \right),
                & \mbox{$x \geq r_2$} \qquad \mbox{(a)}
                \\
\left( 1 - \dfrac{m_0}{x^2} -\kappa\dfrac{x^2 - r_1^2}{x^2}\right),
& \mbox{$r_1\leq x \leq r_2$} \quad \mbox{(b)}
                \\
\left( 1 - \dfrac{m_0}{x^2}\right), & \mbox{$ x \leq r_1$} \qquad
\mbox{(c)}.
\end{cases}
\end{equation}

So, we obtained the lens equation (\ref{eq8}) for the system
consisting of a central mass and a homogeneous annular disc. Beyond the
outer boundary of disc, the images are identical to those from
the point mass equal to the total mass of the system (\ref{eq8}a).
In the hole of the disc, only the central mass $m_0$ (\ref{eq8}c)
influences on the light deflection.  And, finally, the disc itself
forms images with the characteristics depending mainly on the surface
density distribution $\kappa$ (\ref{eq8}b).

We note that in the limiting case $m_0 \rightarrow 0$, $m_{disc} =
1-m_0 \rightarrow 1$ and the equations (\ref{eq8}) transform to the lens
equation for the case of a homogeneous disc obtained by Schneider,
Ehlers, Falco (1999). Since this case has not been studied in
detail, we will consider it briefly in Section 5.

In Fig.1, the images of a grid of sources, appearing due to lensing
by a homogeneous annular disc (Fig.1a) and by a system “central mass +
disc”, are presented.  It is seen from Fig.1a that lensing by
the disc leads to formation of two Einstein rings. In this case, the
images in the hole of the disc remain undistorted, but the images in
the outer region of the disc are distorted as due to lensing by a
point mass $m=1$. The presence of mass in the centre of disc
symmetry can lead to formation of three Einstein rings (Fig.1b).
Thus, the images have the shape as due to lensing by a point mass
$m_0$ in the region inside the third Einstein ring. (The idea  to
represent the lensing images for the grid of Gaussian sources  was
suggested by Victor Vakulik.)

The images in Fig. 1 and all subsequent figures were obtained in the
following way. We divide the selected region in the image plane into
a grid of coordinates $\boldsymbol{x}_i$  with a step $h$. Then, we
calculate the corresponding values of coordinates $\boldsymbol{y}_i$
using the analytical expressions (\ref{eq8}) for the lens system
with the defined parameters ($m_0$, $r_1$, $r_2$). From the obtained array
($\boldsymbol{x}_i$, $\boldsymbol{y}_i$) , we chose such
$\boldsymbol{y}_i$ for which the difference $|\boldsymbol{y}_i -
\boldsymbol{r}_i| < h$, where $\boldsymbol{r}_i$ are the source coordinates. 
So, we find corresponding coordinates of image
$\boldsymbol{x}_i$ by obtained values $\boldsymbol{y}_i$.

\subsection{Radii of the Einstein rings (for the case of a point source)}

The equations (\ref{eq8}) show that with $\boldsymbol{y}=0$ (when
point source, lense centre and observer are aligned) three Einstein
rings can be formed. It is convenient to number the rings in
order of decreasing of their radii (Fig.2). Hereafter,  the ring of
a radius $x_\text{I} = 1$   we call the first (or main) ring (the
solution of the equation (\ref{eq8}a)). The first ring forms if
$x_\text{I}>r_2$ for $r_2<1$. The equation (\ref{eq8}b) for
$\boldsymbol{y}=0$ leads to quadratic equation which is easy to solve
relative to $x$:
\begin{equation}\label{eq9}
x_\text{II} =\sqrt{\left|\frac{\kappa r_1^2 - m_0}{\kappa - 1}\right|} = \sqrt{\left| \frac{\kappa r_2^2 - 1}{\kappa - 1}\right| }
\end{equation}
Thus, the second Einstein ring is the ring with radius $x_\text{II}$
which is always formed on a lensing disc ($r_1 < x_\text{II} < r_2$)
and depends on the disc surface density. It is seen from the
equation (\ref{eq8}c) that in this system the third Einstein ring
appears with the radius $x_{\text{III}}= \sqrt{m_0}$. The third ring
appears in the hole of the disc $x_\text{III} < r_1$ or $r_1 >
\sqrt{m_0}$.

\section{WIDTHS OF THE EINSTEIN RINGS AND ITS BOUNDARIES}

A finite size of a source has its effect on the resulting image. Overlapping or
merging of the rings is possible in this case. Consider this in detail by examining
the boundaries of the Einstein rings for each case separately.

a) The inner and outer boundaries of {\it the main Einstein ring} have the next view:
\begin{equation}\label{eq10}
x_\text{I}^{Out,In} = \pm \frac{r_s}{2} + \sqrt{\left(\frac{r_s}{2}\right)^2 + 1}
\end{equation}
where the signs plus/minus correspond to the outer ($x_\text{I}^{Out}$) and inner($x_\text{I}^{In}$)
boundaries of the ring, and $r_s$ is a sources radius. The expression (\ref{eq10})
is easy to obtain from equation (\ref{eq8}a) with $\boldsymbol{y}=\pm r_s$. The width of the main Einstein ring equals the source radius: $\triangle x_\text{I} = x_\text{I}^{Out} - x_\text{I}^{In} = r_s$.
It is obvious that the main ring can overlap with the lensing disc because the width of the main ring
depends on the radius of the source. The main ring is entirely seen  if $r_2 <x_\text{I}^{In}$.
Otherwise, the main ring can be screened by the disc, either partially ($x_\text{I}^{In}<r_2<x_\text{I}^{Out}$)
or fully ($r_2 \geq x_\text{I}^{Out}$).

In Fig.3, the images of the grid of sources formed by the lensing system “point mass+disc” are presented.
Fig.3a corresponds to the case when the main ring does not overlap with the disc ($x_\text{I}^{In} \approx 0.95 > r_2$)
and all the three Einstein rings are clearly seen.The case of partial overlapping of the main ring with
the lensing disc ($x_\text{I}^{In} < r_2$,  $x_{I}^{Out} \approx 1.05 > r_2$) is presented in Fig.3b.
In addition, the first and the second rings merge in one (wide) Einstein ring for the given parameters of the lensing system.
As a result, we see two Einstein rings.  In Fig.3c, the main ring is fully screened by the disc
($r_2 > x_\text{I}^{Out}$). In this case the second Einstein ring vanishes and the resulting image is only
one ring (the third Einstein ring). Thus, there is a variety of cases resulting in formation of one, two and three Einstein rings.

b) Generalizing the expression (\ref{eq10}) for the case of a point mass $m_0$, we can immediately write down the expressions for the inner and outer boundaries of {\it the third Einstein ring}:
\begin{equation}\label{eq11}
x_\text{III}^{Out,In} = \pm \frac{r_s}{2} + \sqrt{\left(\frac{r_s}{2}\right)^2 + m_0}.
\end{equation}
\begin{figure}\label{Fig4}
\includegraphics[width = 80mm]{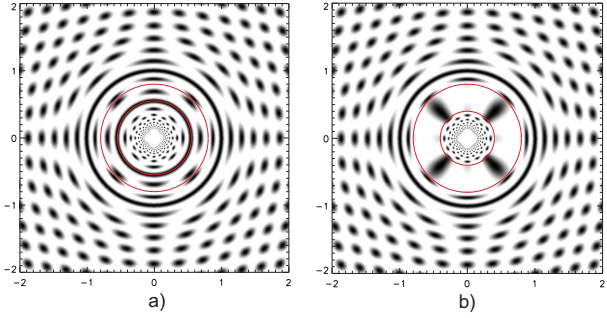}
\caption{Images of the grid of Gaussian sources ($r_s=0.1$) due to gravitational lensing by
a central point mass with $m_0=0.3$ and a homogeneous disc with the inner radius $r_1=0.7$,
for three cases: a) $r_2=0.9$, b)$r_2=0.98$, c)$r_2=1.2$.}
\end{figure}
\begin{figure}\label{Fig5}
\includegraphics[width = 75mm]{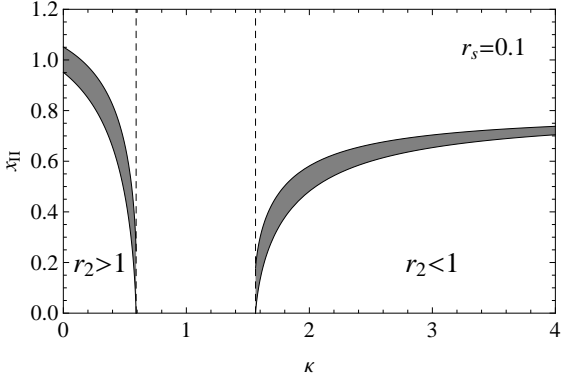}
\caption{Dependence $x_\text{II}^{Out,In}(\kappa)$ for two cases:   $r_2 = 0.8$ and $r_2 = 1.3$. Grey regions show the change of the second Einstein ring width. “Exclusion region”  of the second ring are marked by the dashed lines  and corresponds to $\kappa > \kappa_{cr}$ (for $r_2<1$) and $\kappa < \kappa_{cr}$ ($r_2>1$).}
\end{figure}
We obtain the median radius of the third ring
$r_\text{III}=\sqrt{m_0}$ from (\ref{eq11}) for $r_s=0$. Obviously,
the width of the main and the third rings equals the source
radius $\triangle x_\text{III}=\triangle x_\text{I}=r_s$. There are
also conditions for visibility of the third ring. If the outer
radius of the third ring is less than the disc radius
$x_\text{III}^{Out} < r_1$, the third ring is entirely formed (Fig.
3a). A partial overlapping of the third ring by the disc occurs for
the following conditions : $x_\text{III}^{Out} > r_1$,
$x_\text{III}^{In} < r_1$.

The case of partial overlapping of the third Einstein ring is presented
in Fig.4a.  In this case $x_\text{III}^{In} =0.5$ and
$x_\text{III}^{Out} =0.6$ and the second ring merges with the third
one. The parameters of the lens system are such that the width of
the second ring is almost equal to the width of the third ring.
Thus, it is difficult to distinguish the resulting image
from that in  the case of lensing by two point masses, where two
Einstein rings are also formed (Werner, An \& Evans 2008). In Fig.4b,  the
third ring is completely screened by the disc because the condition
$x_\text{III}^{In} > r_1$ is satisfied. Here, we have the case
similar to that at Fig.3c, i.e. only one of the three possible rings
is formed (the main Einstein ring). The second ring is not formed
for the given surface density of the disc. (We will investigate
conditions for existence of the second ring in the next
section.)

c) As opposed to the main and third rings,  boundaries and  widths
of {\it the second Einstein ring} are more complicated. The equation
(\ref{eq8}b) for $\boldsymbol{y}=\pm r_s$  leads to  the quadratic equation:
\begin{equation}\label{eq12}
(1-\kappa)x^2 \pm r_s x + \kappa r_2^2 - 1=0.
\end{equation}
Solution of this equation in general case has the form
\begin{equation}\label{eq13}
x_\text{II}^{Out,In} = \frac{1}{2(1-\kappa)}
                \left[
                \pm r_s \pm \sqrt{r_s^2 -4(1-\kappa)(\kappa r_2^2 - 1)}
                \right].
\end{equation}
Note that the sign before  the square root is determined by the
parameters of  lensing system. Width of the second ring depends on
the lensing disc density and in most cases (see next section) can be
written as
\begin{equation}\label{eq14}
\triangle x_\text{II} = \frac{r_s}{|\kappa - 1|}.
\end{equation}
Thus the width of the second ring depends not only on the source
radius but also on the disc surface density (see Figs 3a,b, 4b).
Since a source in astrophysical problems is nearly a point one, we
will investigate the conditions for existence of the second Einstein
ring restricting ourselves to the case $r_s \ll 1$.

\section{THE CONDITIONS FOR EXISTENCE OF THE SECOND EINSTEIN RING}
\begin{figure*}\label{Fig6}
\includegraphics[width = 150mm]{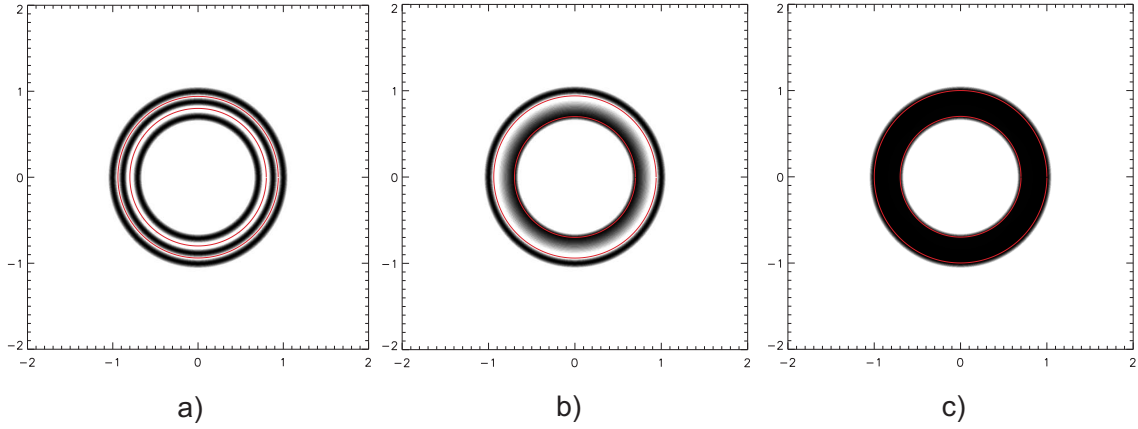}
\caption{Images of a Gaussian source ($r_s=0.1$) formed due to
gravitational lensing by a central mass $m_0=0.5$ and a
homogeneous disc with $r_2<1$ for the following cases: a) $r_1=0.8 >
x_\text{III}^{Out}$, $r2=0.94 <x_\text{I}^{In}$, b) $r_1=0.7 <
x_\text{III}^{Out}$, $r_2=0.94$,  c) $r_1=0.7$, $r_2=1.0$.}
\end{figure*}
\begin{figure*}\label{Fig7}
\includegraphics[width = 150mm]{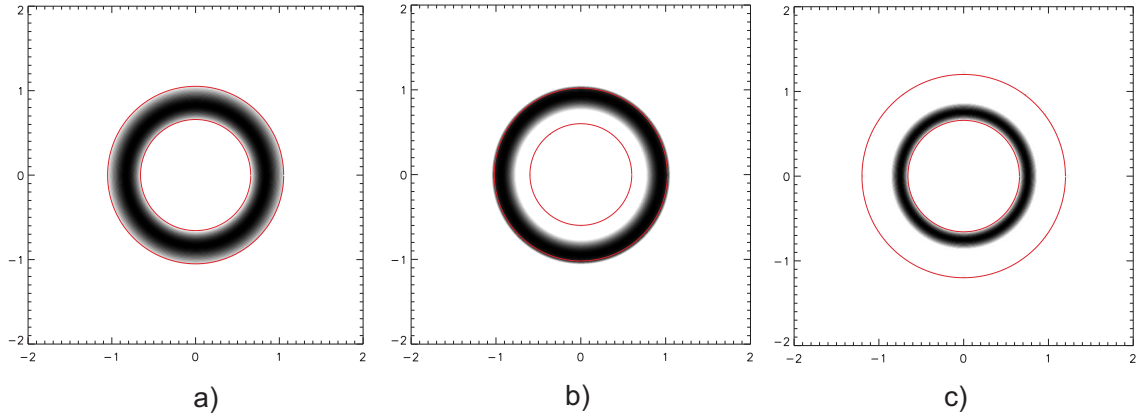}
\caption{Images of a Gaussian source ($r_s=0.1$) formed due to
gravitational lensing by the central mass $m_0=0.5$ and the
homogeneous disc with $r_2<1$ for the next cases: a) $r_1=0.8 >
x_\text{III}^{Out}$, $r2=0.94 <x_\text{I}^{In}$, b) $r_1=0.7 <
x_\text{III}^{Out}$, $r_2=0.94$,  c) $r_1=0.7$, $r_2=1.0$.}
\end{figure*}
It is seen from equation (\ref{eq13}) that for $r_s \ll 1$, the expression for
the inner and outer boundaries of a ring can be written as
\begin{equation}\label{eq15}
x_\text{II}^{Out,In} = \frac{1}{2|\kappa-1|}
                \left[
                \sqrt{r_s^2 +4|(\kappa -1)(\kappa r_2^2 - 1)|} \pm r_s
                \right]
\end{equation}
on conditions that a)$\kappa >1$, $\kappa r_2^2 >1$ or b)$\kappa
<1$, $\kappa r_2^2 <1$. These conditions correspond to 
existence of the real values of the expression (\ref{eq15}). It is convenient to analyse the solutions,  considering two different cases ($r_2 < 1$ and $r_2
> 1$) and introducing the next notation $\kappa_{cr} = 1/r_2^2$ or $\kappa_{cr} = m_0/r_1^2$.
Then {\it the necessary conditions for existence of the second
Einstein ring} are
\begin{equation}\label{eq16}
\begin{matrix}
\mbox{for} & r_2>1 & \mbox{it is necessary that} & \kappa
<\kappa_{cr} & (\kappa <1) & a) \\ \mbox{for} & r_2<1 & \mbox{it is
necessary that} & \kappa > \kappa_{cr} & (\kappa >1)& b).
\end{matrix}
\end{equation}
The physical sense of these conditions is a 'competition' of gravitational influence between gravitational effects from the central mass and the disc.
In Fig.5, dependences of radius and width of the second ring on
the surface density for chosen values of the outer radius of the
disc $r_2<1$ and  $r_2>1$ are shown.  The region of values $\kappa$  for
which the second ring does not exist we call the 'exclusion region'.
For $\kappa \gg \kappa_{cr} \gg 1$  ($r_2 < 1$), the width of the
second ring is decreased but its radius tends to $r_2$
asymptotically. This case corresponds to a limiting passage to
an infinitely thin ring which we will consider in the next section. On
the other hand, for $\kappa \rightarrow 0$ ($r_2 > 1$) the medium
radius of the second ring $x_\text{II}\rightarrow 1 = x_\text{I}$
and width $\triangle x_\text{II} \rightarrow r_s$. This case
corresponds to the limiting passage to lensing by only the central mass,
$m_0 \rightarrow 1$.

The necessary conditions (\ref{eq16}) are not sufficient ones,
because for real values of the inner and outer radii of the second ring
they can go beyond the domain of applicability (\ref{eq8}b), i.e.
abroad the lensing disc. For example, it is seen from (\ref{eq15})
that for $\kappa \rightarrow 1/r_2^2 \pm 0$ the outer radius  $x_\text{II}^{Out}
\rightarrow \triangle x_\text{II}$ and $x_\text{II}^{In}\rightarrow
0$. However if $x_\text{II}<r_1$ the second ring is not formed. The
sufficient conditions for existence of the second ring are
$x_\text{II}^{Out}\leq r_2$ and $x_\text{II}^{In}\geq r_1$. If these two
cases are realized simultaneously, the second Einstein ring is seen
entirely. In other case, only a part of the second ring is formed
due to screening by the disc.

The conditions of the full visibility of the second Einstein ring
have the following form: $x_\text{II}^{Out}\leq r_2$ and
$x_\text{II}^{In}\geq r_1$. These conditions can be expressed using
a minimal set of parameters ($r_s$ and $m_0$):
\begin{equation}\label{eq17}
\begin{matrix}
\mbox{for} & r_2 \geq 1  & r_1 \leq x_\text{III}^{In} & \mbox{and} & r_2 \geq x_\text{I}^{Out} & a) \\
\mbox{for} & r_2 \leq 1 & r_1 \geq x_\text{III}^{Out}& b)
\end{matrix}
\end{equation}
If conditions (\ref{eq16}b)  and  (\ref{eq17}b)  are satisfied (for
$r_2 \leq 1$) than the second ring always exist. In this
case, the third ring is entirely seen but the first ring can be
partly screened by the disc for $r_2 > x_\text{I}^{In}$. Thus, if we
define the radius of the source $r_s$ and the value of the central
mass $m_0$, we can obtain the values  $r_1$, $r_2$, for which three,
two or one Einstein ring(s) are formed. In Figs 6 and 7,
we presented the results of simulations for $m_0 = 0.5$ and
$r_s=0.1$. In this case, three Einstein rings can exist for the given
values of the disc radii (Fig.6a). We note that if the central mass
increases (for fixed $r_1$, $r_2$), the third ring is screened by
the disc because $x_\text{III}=\sqrt{m_0}$. In this case, the radius
of the second ring decreases and gradually disappears because
$x_\text{II}^{Out,In}$ extends beyond the disc (the condition
(\ref{eq17}b)). In Fig.6b we can see two Einstein rings. The inner
ring is a result of merging the third and partly the second
rings. Fig. 6c shows a special case: the surface density in the disc
$\kappa =0.98 \rightarrow 1$. In this case the width of the second
ring $\triangle x_\text{II} \rightarrow \infty$ (see (\ref{eq14})).
As a result, a single Einstein ring is formed with the width equal
to the disc width, and the intensity in each point of the ring equals
the source intensity at its maximum (the central point of the
Gaussian source).
\begin{figure*}\label{Fig8}
\includegraphics[width = 150mm]{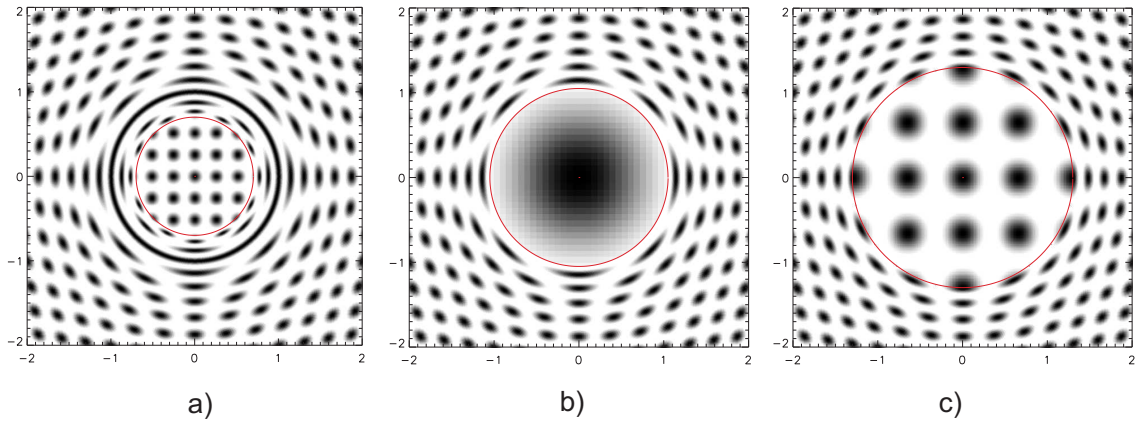}
\caption{Images of the grid of sources  ($r_s=0.1$) formed due to
gravitational lensing by a homogeneous disc ($m_{disc} =1$ and
$r_1=0$)  for the cases: a) $r=0.7$, b)$r=1.05= x_\text{I}^{Out}$,
c) $r=1.3$.}
\end{figure*}

If the outer radius of the disc $r_2 >1$, then the image is a single
Einstein ring,  but with different width due to the merging together
the first and the second rings or the  second and the third rings
(Fig.7). Fig.7a shows the limiting case of the conditions
(\ref{eq17}a) $r_1 = x_\text{III}^{In}$ and $r_2 =
x_\text{I}^{Out}$, for which the image is a wide (second) ring
entirely filling the disc area entirely. In Fig.7b, the first ring is
screened partly by the disc and merges  with the second ring. In
this case, the conditions (\ref{eq17}a) are not satisfied and the
outer radius of the second ring $x_\text{II}^{Out}$ extends beyond
the disc, i.e. the second ring is seen partly.  In Fig.7c, the second
ring contacts the inner boundary of the disc and is visible
completely. In this case, the radius of the
second ring $x_\text{II}$ increases up to unity with decreasing $m_0$. And vice versa, as the outer radius of the disc (for a given fixed
$m_0$) increases, the radius $x_\text{II}$ decreases and the second ring
is gradually disappearing. Vanishing of the second ring is
accompanied by appearance of the third Einstein ring. Such
dependence of the second ring radius on the parameters of the
lensing system can be understood by considering the limiting cases. We
will discuss this in the next section.

Note that in addition to the solution (\ref{eq15}) examined above,
there is a solution of the form
\begin{equation}\label{eq18}
x_\text{II}^{Out,In} = \frac{1}{2|\kappa -1|}
                \left[
                r_s \pm \sqrt{r_s^2 -4|(\kappa-1)(\kappa r_2^2 - 1)|}
                \right].
\end{equation}
This solution is valid for a source with a large radius, so we will
not dwell on it. We note only that this solution falls into the
range of values for a surface density $1 \leq \kappa \leq
\kappa_{cr}$ for $r_2 \leq 1$ and $\kappa_{cr} \leq \kappa \leq 1$
for $r_2 \geq 1$. For a small source these regions are very narrow 
and abut to the values $\kappa\rightarrow 1$ and/or $\kappa
\rightarrow \kappa_{cr}$. For a source with a large value of $r_s$
these regions can work but only in a narrow range of parameters of
the lensing system.

We can predict without simulation the configuration of Einstein rings if a central mass in our system is a black hole. In this case, all the features of the Einstein rings, discovered above, will be the same but with additional (relativistic) rings in the nearest vicinity of black hole.

\section{LIMITING CASES}

The solution, obtained for the lensing system "a homogeneous disc
with a hole + central mass", is a generalization of the known cases
of more simple systems. Let us consider them briefly because  it
will be useful for understanding the rings formation in the given
system for different lens parameters.

a) Passage to the limit of a point mass  $m_0 \rightarrow 1$
($m_{disc}\rightarrow 0$, $\kappa\rightarrow 0$). It is easy to see
that $x_\text{II}\rightarrow x_\text{I}=1$ and $\triangle
x_\text{II} \rightarrow r_s$, i.e. the second ring degenerates to
the main Einstein ring. Thus, if the disc mass is decreasing while keeping its
surface area, the radius of the second ring tends to the radius of
the main Einstein ring.

b) Passage to the limit of a wide disc $r_2 \gg 1$ also corresponds
to $\kappa \rightarrow 0$. Since the central mass has a fixed value
$m_0 <1$, we have  $x_\text{II}^{Out,In} \rightarrow (\sqrt{r_s^2 +
4|\kappa r_1^2-m_0|}\pm r_s)/2 \rightarrow x_\text{III}^{Out,In}$
and the second ring degenerates to the third Einstein ring. Note
that increasing of the inner radius of the disc results in 
a decrease of the second ring radius. If the third ring is
screened by the disc, we can see only the second ring (Fig.7c).  On
the other hand, if the value of the central mass decreases, the
second ring disappears, because $r_\text{II}\rightarrow
r_\text{III}=\sqrt{m_0}$. In this case disappearance of the
second ring is accompanied by appearance of the third ring.

c) Passage to the limit of an infinitely thin ring $r_1 \rightarrow
r_2 \rightarrow r$ corresponds to $\kappa \rightarrow \infty$. It is
seen from equations (\ref{eq9}) and (\ref{eq14}) that in this case  the
lens equation (\ref{eq8}) has the form
\begin{equation}\label{eq19}
\boldsymbol{y} = \boldsymbol{x} \times
\begin{cases}
\left( 1 - \dfrac{m_0}{x^2} \right),
                & \mbox{$x < r$} \qquad \mbox{(a)}
                \\
\left( 1 - \dfrac{1}{x^2}\right), & \mbox{$ x > r$} \qquad \mbox{(b)}.
\end{cases}
\end{equation}
If  $r<1$,  the first and the third rings exist simultaneously that
is possible only in this limiting case. In the system a disc and a
central mass, the second ring disappears always in a pair with the
first or third ring. For $m_0=0$, the solution corresponds to
a well-known solution for lensing by an infinitely thin ring
(Zakharov 1997). In this case, there is a single Einstein ring
and an undistorted image of the source in the centre of the system.
\begin{figure*}\label{Fig9}
\includegraphics[width = 150mm]{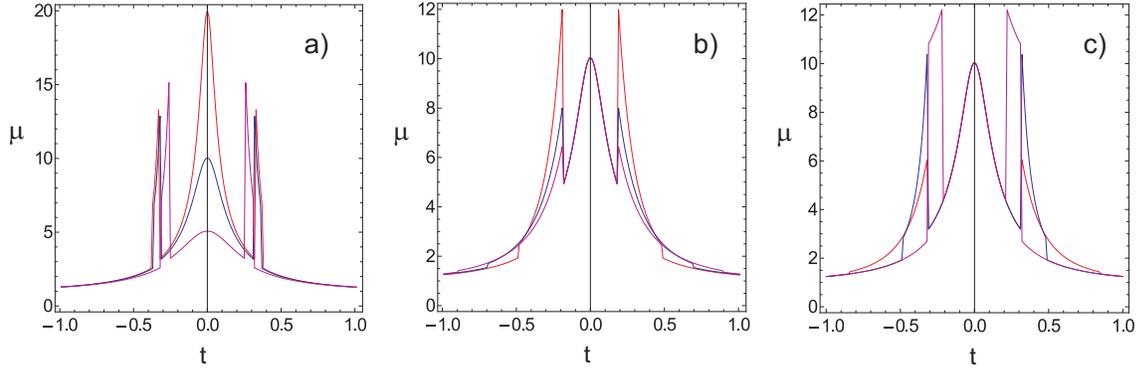}
\caption{Dependence of magnification factor on a source position a) at the disc with $r_1=0.67$,
$r_2=0.85$ and the central mass with $m_0=0.7$ for $r_0=0.05$
(orange), 0.1 (blue), 0.2 (magenta); b) at the disc with $r_1=0.5$,
$r_2=0.9$ for the different values of the central mass: $m_0=0.5$
(orange); $m_0=0.6$ (blue); $m_0=0.7$ (magenta); c)$m_0=0.5$,
$r_2=0.85$, $r_1=0.4$ (orange), $r_1=0.5$ (blue), $r_1=0.6$
(magenta).}
\end{figure*}

d) Passage to the limit of a solid disc (without hole) $r_1
\rightarrow 0$ lead to
\begin{equation}\label{eq20}
\boldsymbol{y} = \boldsymbol{x} \times
\begin{cases}
\left( 1 - \dfrac{1}{r^2} \right),
                & \mbox{$|x| < r$} \qquad \mbox{(a)}
                \\
\left( 1 - \dfrac{1}{x^2}\right), & \mbox{$|x| > r$} \qquad
\mbox{(b)}
\end{cases}
\end{equation}
where we denote $r_2=r$ for simplicity.  This is a well-known solution
( Schneider, Ehlers \& Falco 1993; Zakharov 1997),
but it  seems useful to consider the images
for some specific  cases. It is seen from equation (\ref{eq20}a) that the
radius of an image formed by the disc increases or decreases 
depending on the disc radius: $r_{im}=r_s |1 - 1/r^2|^{-1}$. For
$r<1/\sqrt{2}$ the image size is less than the source size (Fig.8a),
but for $1/\sqrt{2} <r <1$ the image of the source increases and
tends to infinity for $r \rightarrow 1 \pm 0$. Fig. 8b presents the
case where the image is magnified by a factor of about $11$ and almost
completely fills the disc. Further increasing of the disc radius
leads to decreasing of the image size (Fig.8c), and $r_{im} \rightarrow r_s$
for $r\rightarrow \infty$ . Similar effect appears in the
considered system 'disc + central mass'. Increasing of the image
formed by disc (for shifted source) can lead to significant effects
on magnification, which we discuss in the next section.

\section{MAGNIFICATION FACTOR}

Consider a dependence of magnification factor on a source position for the case of a point source.
Critical curves are defined by the condition $\det A=0$  and the
magnification factor is $\mu =1/\det A$, where the matrix determinant $\det
A = |\partial y_i/\partial x_j|$. The lens equation (\ref{eq8})
gives
\begin{equation}\label{eq21}
\det A =
\begin{cases}
 1 - \dfrac{1}{x^4} ,
                & \mbox{$x \geq r_2$} \qquad \mbox{(a)}
                \\
(1 - \kappa)^2 -\left(\dfrac{\kappa r_2^2 - 1}{x^2} \right)^2,
                & \mbox{$r_1 \leq x \leq r_2$} \qquad \mbox{(b)}
                \\
1 - \dfrac{m_0^2}{x^4}, & \mbox{$x \leq r_1$} \qquad
\mbox{(c)}.
\end{cases}
\end{equation}
It is easy to see from (\ref{eq21}) that the critical curves are
circles with radii equal to the radii of Einstein rings:
$x_\text{I}^{cr} = x_\text{I}=1$, $x_\text{II}^{cr}=x_\text{II} =
\sqrt{|(\kappa r_2^2 -1)/(\kappa -1)|}$, $x_\text{III}^{cr} =
x_\text{III} = \sqrt{m_0}$. The caustic (including all three
regions) is a point with coordinates ($0,0$). The magnification factor for
each of the three regions has the form
\begin{equation}\label{eq22}
\begin{aligned}
\mu_\text{I}^\pm =\dfrac{1}{4}
    \dfrac{
        \left(
        y \pm \sqrt{y^2 + 4}
        \right)^2}
    {y\sqrt{y^2 + 4}} \qquad \mbox{(a)}
                \\
\mu_\text{II}^\pm =\dfrac{1}{4(\kappa - 1)^2}
    \dfrac{
        \left(
                y \pm \sqrt{y^2 + 4(\kappa-1)(\kappa r_2^2 - 1)}
        \right)^2}
        {y\sqrt{y^2 + 4(\kappa-1)(\kappa r_2^2 - 1)}} \qquad \mbox{(b)}
                \\
\mu_\text{III}^\pm =\dfrac{1}{4}
    \dfrac{\left(
            y \pm \sqrt{y^2 + 4m_0}
            \right)^2}
            {y\sqrt{y^2 + 4m_0}} \qquad \mbox{(c)}
\end{aligned}
\end{equation}
The sign '+' corresponds to an image of positive parity, the sign
'-' corresponds to an image of negative parity. As one would expect,
the magnifications for the first (\ref{eq22}a) and third
(\ref{eq22}c) regions present the magnification for the point lens
with a mass equal to $1$ and $m_0$, respectively. Note that the
magnification factor (\ref{eq22}b) can be obtained by using the following
substitution in the expression (\ref{eq22}a): in the denominator $4
\rightarrow 4(\kappa - 1)^2$ and under the square root sign $4
\rightarrow 4(\kappa - 1)(\kappa r_2^2 - 1)^2$. The total
magnification taking account of all regions is $\mu = \mu_\text{I} +
\mu_\text{II} + \mu_\text{III}$.

Let us consider a source which moves in a straight line at a certain
distance from the centre of a gravitational lens. It is convenient
to parametrize $y$ in the follow form
\[
y = \sqrt{r_0^2 + t^2}
\]
where $r_0$ is the minimal distance the source approaches the
origin of a coordinate system, $t$ is the source position relative 
to the lens centre (for $t=0$ $y=r_0$).  The
magnification curves are obtained in the follow way. We calculate values $y_\pm$ for
a given $t$ and then calculate the values $x_\pm$ from the lens equation (\ref{eq8}) for different regions, and the corresponding magnifications from the
formula (\ref{eq22}).

In Fig.9, we show various examples of dependence of magnification factor on a source position (name them “the magnification curves” for convenience). Three well-defined peaks are seen in all the three figures.
There are differences
in the peak shapes from the disc and the central mass. It is seen
from  Fig.9a that contribution to the magnification due to
lensing by the disc is becoming predominant with an increase of $r_0$. On the
other hand, the highest values of magnification are achieved at the
minimum value of the central mass (Fig.9b).  Fig.9c shows that
different widths of the disc result in different shapes of the magnification
curves.

\section{Conclusion}
Analysis of the Einstein rings arising due to lensing by a disc and a central mass has shown a diversity of possible cases.
First, this is formation of three Einstein rings. Assuming that in most of astrophysical objects $m_0 \geq m_{disc}$,
we obtain rough limit on the radius of the disc for which three Einstein rings may exist: $r_1 \geq 0.7$ and $r_2 <1$
at surface density $\kappa > \kappa_{cr}$. The more common case is two Einstein rings which arise in a wide range
of parameters of the lensing system. It is shown that in this case the surface brightness of one of the rings is dependent
on the surface density of the disc and can decrease to the inner or outer boundary of the disc. Formally, knowing
the brightness distribution in a ring, we can restore the density distribution in the disc. The most frequently encountered  case is
one Einstein ring. We have shown that the resulting image can be indistinguishable from the main Einstein ring or,
conversely, it may be of a substantial width, thus filling the disc completely. In fact, the ring distribution of matter
in the lens may have the same effect as a point mass. If the annular lens is not detected in the optical light
(for example dark matter), one can make wrong conclusions about the mass and its distribution in the lens,
as well as about the distance to it. In contrast, the observation of a wide bright Einstein ring can be a criterion
for  identification of objects with the ring structures. Note also that the source image can be magnified while passing
across the disc (in projection on the lens plane). This leads to a significant rise in the magnification curve, which may also
be a criterion for detection of such lenses.

For real objects accounting for
inhomogeneous density distribution in the annular lens and for possible
inclination of the lens plane to the plane of the sky are needed. All these may be
the next step in investigations of the gravitational lensing
effects by objects having the ring (toroidal)
structures.

\section*{Acknowledgements}
The idea of this work arose thanks to the numerical simulations made by Victor Vakulik for the ring-like lens consisting of $N$-point masses.

We thank V.M.Shulga for support of our investigation. We are grateful to V.S. Tsvetkova, V.M. Kontorovich, 
P.P. Petrov, D.G. Stankevich and the Kharkov grav.lensing group for helpful discussions and remarks. 
We thank the team of Institute Henri Poincare (Paris) and
Centre Emile Borel for their hospitality during our stay and financial support.
EYB is grateful to The European Astronomical Society (EAS) for the travel grant, 
which allowed us to present the results of this work at the EWASS-2014 (Geneva).

\end{document}